\documentclass{ws-ijmpc}

\begin{document}

\markboth{Dami\'an H. Zanette} {Analytical approach to bit-string
models of language evolution}

\catchline{}{}{}{}{}
%%%%%%%%%%%%%%%%%%%%%%%%%%%%%%%%%%%%%%%%%%%%%%%%%%%%%%%%%%%%%%%%%%%%%

\title{ANALYTICAL APPROACH TO BIT-STRING MODELS OF LANGUAGE EVOLUTION}

\author{DAMI\'AN H. ZANETTE}

\address{Consejo Nacional de Investigaciones Cient\'{\i}ficas y
T\'ecnicas\\
Centro At\'omico Bariloche and Instituto Balseiro\\
8400 San Carlos de Bariloche, R\'{\i}o Negro, Argentina.\\
zanette@cab.cnea.gov.ar}

\maketitle

\begin{history}
\received{Day Month Year}
\revised{Day Month Year}
\end{history}

\begin{abstract}
A formulation of bit-string models of language evolution, based on
differential equations for the population speaking each language, is
introduced and preliminarily studied. Connections with replicator
dynamics and diffusion processes are pointed out. The stability of
the dominance state, where most of the population speaks a single
language, is analyzed within a mean-field-like approximation, while
the homogeneous state, where the population is evenly distributed
among languages, can be exactly studied. This analysis discloses the
existence of a bistability region, where dominance coexists with
homogeneity as possible asymptotic states. Numerical resolution of
the differential system validates these findings. \keywords{Language
evolution; replicator dynamics; diffusion.}
\end{abstract}

\ccode{PACS Nos.: 87.23.Ge, 87.23.Kg, 05.45.-a}

\section{Introduction}

The dynamics of language has recently been identified as a rich
field for interdisciplinary application of statistical techniques
traditionally associated with mathematics and physics. A host of
analytical and numerical models have been proposed, aimed at
reproducing --more or less quantitatively-- several aspects of
language as a dynamical sociocultural
entity.\cite{nowak,cancho,strog,kosm,schulze,schw,viviane,tes,t1}
Among them, models of language evolution focus on the joint
processes of mutation of linguistic features and of language
acquisition, switching, and adoption by human
populations.\cite{schulze,viviane,t1}

Bit-string models of language evolution, prototyped by Schulze's
model,\cite{schulze} conceive that a language is completely
characterized by a sequence of dichotomic properties. Each of them
represents, for instance, whether the language in question possesses
certain grammatical property or not. The sequence is naturally
represented as a string of binary variables (bits), whose length is
the number of yes/no questions which fully identify a language.
Emphasis in the study of bit-string language models has been put on
agent-based numerical simulations,\cite{schulze,life,sol,wich,oli0}
addressing possible explanations for the distribution of population
among languages and the abundance of language families. The language
spoken by an individual can mutate at a certain rate, and each
individual can also abandon his/her language and adopt a different
one copied from a randomly chosen member of the population. Upon
variation of the mutation rate, simulations show a sharp transition
between a state of dominance, where most of the population speaks
the same language, and a situation where the population is
homogeneously distributed among languages.\cite{tes,wich}

In this paper, I introduce a formulation of bit-string language
models based on differential equations for the population fraction
speaking each language. The evolution turns out to be a combination
of replicator dynamics, representing language switching, and
diffusion in the bit-string (hypercubic) space, representing
mutation. While the stability of the homogeneous state can be
exactly analyzed, the study of the dominance state requires a
mean-field-like approximation. The results, nevertheless, are in
very good agreement with numerical resolution of the differential
equations and with agent-based simulations. They predict a region of
bistability, where dominance and homogeneity coexist as possible
asymptotic states of the system, and the disappearance of the
dominance state through a tangent bifurcation. The present
formulation provides an alternative to simulations to trace language
progress and regression in the framework of bit-string models.

\section{Analytical formulation of bit-string language models}

As advanced in the Introduction, in bit-string models an individual
language is represented by a string of $L$ binary variables (bits),
each of them adopting one of two possible values, say, $0$ and $1$.
The total number of possible languages is $N=2^L$. Evolution is
driven by two dynamical rules. First, the language of an individual
can mutate with a certain probability per time unit. Each mutation
event consists of a change in a single, randomly selected bit of the
individual's language, from $0$ to $1$ or {\it vice versa}. Second,
an individual can give up his/her language and adopt a new one. To
account for the preference for more widespread languages, the new
one is chosen by selecting an individual at random from the whole
population and adopting his/her language. In this way, the
probability of switching to a given language (say, to language $i$)
equals the fraction of the population speaking $i$, denoted by
$x_i$. Moreover, the probability of abandoning the original language
(say, language $j$) is weighted by a monotonically decreasing
function $u(x_j)$ of the fraction of the population speaking $j$. A
language is thus less likely to be given up if it is spoken by a
large fraction of the total population. Previous numerical analyses
of bit-string models  have considered the cases $u(x)=1-x^2$ and
$u(x)=(1-x)^2$, both of which satisfy $u(0)=1$ and
$u(1)=0$.\cite{schulze,life,wich}

Bit-string models are added with population dynamics, for instance,
asexual reproduction with births at a given rate, and deaths with a
mortality rate proportional to the population size. This insures
that --after a transient, and up to small fluctuations-- the total
population remains constant in time.  A newborn inherits the
language of the parent, except possibly for mutation or switching to
a new language, following essentially the same rules as described
above. The effect of births on language evolution, thus, is
effectively the same as mutation and switching by ``adult''
individuals.

In its original formulation,\cite{schulze} Schulze's model considers
that the features of a given (``superior'') language may be
preferred to those of any other language. During mutation events,
with a certain probability, a bit copies the value of the homologous
bit in the ``superior'' language, enhancing the possibility that
this language is eventually spoken by most of the population. Here,
I disregard this mechanism and assume that all languages are
intrinsically equivalent.

The dynamical rules of bit-string language models can be readily
expressed as differential equations for the fractions $x_i(t)$
($i=1,\dots,N$) of the population speaking each language, which
satisfy the normalization condition
\begin{equation} \label{norm}
\sum_{i=1}^N x_i(t) =1
\end{equation}
at all times. For the sake of simplicity, I will assume that the
total population remains exactly constant in time, incorporating
newborn language's mutation and switching to the respective overall
mechanisms.

The contribution of mutation to the time variation of $x_i$ consists
of a gain term proportional to $L^{-1}\sum_j w_{ij} x_j$, where
$w_{ij}=1$ if languages $i$ and $j$ differ by just one bit, and
$w_{ij}=0$ otherwise. The sum runs over all the languages. The gain
term represents the fraction of the population speaking any language
$j$ which mutates to language $i$ in a time unit. The factor
$L^{-1}$ accounts for the fact that a mutation event can lead, with
identical probability, from each language $j$ to $L$ different
languages. Correspondingly, there is a loss term proportional to
$L^{-1} x_i \sum_j w_{ji}$, which represents mutations of language
$i$ to other languages. In turn, the gain term associated to the
mechanism of language switching is proportional to $x_i \sum_j x_j
u(x_j)$, because the probability of abandoning  $j$ is $u(x_j)$, and
the probability of adopting $i$ is $x_i$. The corresponding loss
term is proportional to $x_i u(x_i) \sum_j x_j$.

If $\mu$ and $\rho$ are, respectively, the rates of mutation and
switching events, the evolution of the fraction $x_i$ ($i=1,2,\dots
, N$) is given by
\begin{equation} \label{fund}
\dot x_i = \rho x_i \left( \sum_{j=1}^N x_j u(x_j) -u(x_i) \right)+
\frac{\mu}{L} \left( \sum_{j=1}^N w_{ij} x_j -x_i \sum_{j=1}^N
w_{ji} \right).
\end{equation}
Summing these equations over the index $i$ yields $\sum_i \dot
x_i=0$, which insures that Eq. (\ref{norm}) holds if the initial
fraction satisfy normalization. The functions $u(x_i)$ make the
equations nonlinear, so that little can be expected from trying to
solve them exactly.

It is worthwhile mentioning that a joint dynamical description of
mutation and population growth --here corresponding to switching to
widespread languages-- has been proposed for the process of language
learning and the evolution of universal grammar.\cite{nsc,nnat}
Several exact results are known for the relevant
equations,\cite{koma} though they depend crucially on the fact that
the functions which play a role analogous to that of $u(x)$ are
linear on the population fractions.

Also, it is important to realize that the representation of
languages as bit strings is essentially irrelevant to the form of
the evolution equation (\ref{fund}). As long as each language is
interpreted as a kind of {\it state} to which a certain fraction of
the population is assigned at each time, the specific way in which
such states are individualized does not play a role in the equation.
At most, it is necessary that those states are suitably labeled, in
such a way as to discern the languages between which population can
be transferred due to mutations, i.e. as to fix de coefficients
$w_{ij}$. Otherwise, the mathematical form of the equations will be
the same irrespectively of the individual characterization of
languages. Though, for clarity, I discuss Eq. (\ref{fund}) with
reference to bit-string models, it is important to bear in mind that
most results will also hold for other models driven by the
mechanisms of mutation and switching.

The contributions of mutation and switching to the evolution of
$x_i$ have opposite, competing effects. While mutation spreads
individuals over different languages, switching tends to concentrate
the population on languages with an already large number of
speakers. It is useful to begin with a separate analysis of the two
mechanisms, which are closely related to well-understood processes
in other areas of science. Then, one can proceed to show that the
combination of the two competing mechanisms gives rise to a critical
transition between language diversity and dominance, as effectively
observed to occur in numerical simulations of bit-string
models.\cite{tes,wich}

\section{Connection with diffusion and with replicator dynamics}

\subsection{Without switching: $\rho=0$} \label{diff}

In order to isolate from each other the effects of mutation and
switching, let us first disregard the latter, setting $\rho=0$ in
Eq. (\ref{fund}). Taking into account that $w_{ij} = w_{ji}$, the
dynamics without switching is governed by the equations
\begin{equation} \label{dif1}
\dot x_i = \frac{\mu}{L} \sum_{j=1}^N w_{ij} (x_j-x_i),
\end{equation}
$i=1,2,\dots,N$. This expression emphasizes the fact that mutation
is a form of linear diffusion, with population transfer between
languages at a rate proportional to the population difference
$x_j-x_i$. The diffusion process takes place on the set of bit
strings of length $L$, which can be assimilated to the vertices of
the $L$-dimensional hypercube of unitary side. Mutation events
transfer population between languages which differ in a single bit,
i.e. between hypercube vertices at Hamming distance $d=1$. Each
vertex has $L$ neighbour vertices at $d=1$.

Since $w_{ij}=1$ if the Hamming distance between $i$ and $j$ equals
one, and $w_{ij}=0$ otherwise, one has $\sum_j w_{ij}=L$. Therefore,
introducing the hypercube adjacency matrix ${\cal W} \equiv \{
w_{ij}\}$,  Eq. (\ref{dif1}) can be rewritten as
\begin{equation} \label{dif2}
\dot {\bf x} = -\mu {\bf x} +\frac{\mu}{L} {\cal W} {\bf x} =
\frac{\mu}{L} {\cal D} {\bf x},
\end{equation}
with ${\bf x}\equiv (x_1,x_2,\dots,x_N)$ and ${\cal D}={\cal W}- L
{\cal I}$, where ${\cal I}$ is the identity matrix. The solution to
the linear equation (\ref{dif2}) can be found by standard methods,
in terms of the eigenvalues $\lambda_k$   and eigenvectors ${\bf
e}_k$ ($k=1,\dots,N$) of the matrix $\cal D$.

One of the eigenvalues of $\cal D$, say $\lambda_1$, equals zero,
and all the remaining eigenvalues are negative. Consequently, the
eigenvector corresponding to $\lambda_1$, ${\bf e}_1=(1,1,\dots,1)$,
yields the equilibrium solution which, due to the normalization
condition (\ref{norm}), is $x_i^*  = N^{-1}$ for all $i$. In other
words, starting from any initial condition, the system
asymptotically reaches a homogeneous state. In the context of the
original problem, in the absence of language switching, mutation
leads for long times to a state where the population is
homogeneously distributed among all possible languages.

\subsection{Without mutation: $\mu=0$} \label{repl}

If, now, $\mu=0$ so that mutation is absent, Eq. (\ref{fund})
becomes a member of a well-studied class of nonlinear equations,
generally written as
\begin{equation} \label{replic}
\dot x_i = x_i f_i({\bf x}) - x_i \sum_{j=1}^N x_j f_j({\bf x}),
\end{equation}
$i=1,2,\dots,N$, with ${\bf x}= (x_1,x_2,\dots,x_N)$. Equations
(\ref{replic}) define the so-called replicator
dynamics.\cite{schuster,sigmund} They describe the evolution of a
system of interacting species in terms of the fractions $x_i=n_i/n$,
where $n_i$ is the number of individuals of species $i$ and
$n=\sum_i n_i$ is the total population in the system. The function
$f_i ({\bf x})$ is the reproductive rate, or (Fisherian
\cite{fisher}) fitness, of species $i$. It is assumed to depend on
the population fractions of all other species. The replicator
equations are at the basis of the mathematical description of
evolutionary dynamics.\cite{kimura,volken}

The first term in the right-hand side of Eq. (\ref{fund}) is
obtained from (\ref{replic}) for $f_i({\bf x}) = -\rho u(x_i)$. In
bit-string language models, therefore, the ``fitness'' to be
assigned to each language is a given function of the corresponding
population fraction, the same for all languages, and does not depend
on the populations of other languages. The negative proportionality
between $u$ and $f_i$ points out that the function $u(x)$ defines a
kind of  ``unfitness.'' In fact, it measures the probability with
which a language is abandoned for adoption of another one.

Though the general solution to the replicator equations is not
known, some generic mathematical properties make it possible to
characterize aspects of the evolution and of its asymptotic state.
These properties can be straightforwardly translated to Eq.
(\ref{fund}) when $\mu=0$. First, the ratio of the populations
fractions of any two languages $i$ and $j$ satisfy the ``quotient
rule'' \cite{sigmund}
\begin{equation}
\frac{d}{dt} \left( \frac{x_i}{x_j} \right) =\rho \frac{x_i}{x_j}
[u(x_j)-u(x_i)],
\end{equation}
which depends on $x_i$ and $x_j$ only. It can be formally
integrated, yielding
\begin{equation} \label{quot}
\frac{x_i(t)}{x_j(t)}= \frac{x_i(0)}{x_j(0)} \exp \left( \rho
\int_0^t [u(x_j)-u(x_i)] dt' \right) .
\end{equation}
Suppose now that the initial population of language $i$ is larger
than that of $j$, i.e. $x_i(0) > x_j(0)$. Since $u(x)$ is a
decreasing function of $x$, the initial value of $u(x_j)-u(x_i)$ is
positive. This implies that the ratio $x_i/x_j$ will grow, making
the difference between the two populations larger. This kind of
feedback effect will be enhanced as time elapses. If, on the other
hand, $x_i(0) < x_j(0)$, the ratio $x_i/x_j$ will monotonically
decrease along the evolution. Since the normalization condition
(\ref{norm}) holds, this result suggests that, generically, the
language which initially has the largest population fraction, say
language $i$, will asymptotically accumulate the whole population,
$x_i (t) \to 1$ as $t \to \infty$, while the fractions of all the
other languages will asymptotically vanish. The only exception to
this behaviour happens if two or more languages have exactly the
same initial population. For two of these languages, in fact,
$u(x_j)-u(x_i)=0$, and $x_i$ and $x_j$ remain identical at all
times. If their initial populations are larger than that of any
other language in the system, they will asymptotically share the
whole population in equal parts, and the remaining populations will
vanish.

The same results arise from a global stability analysis of Eq.
(\ref{fund}), based on the fact that, for $\mu=0$, it admits a
Lyapunov functional,\cite{sigmund}
\begin{equation}
U({\bf x}) = \sum_{i=1}^N \int_0^{x_i} u(x) dx.
\end{equation}
Its time derivative is
\begin{equation}
\dot U = \sum_{i=1}^N u(x_i) \dot x_i =   \rho \left( \langle u
\rangle^2 - \langle u^2 \rangle \right)
 \le 0,
\end{equation}
with $\langle u \rangle = \sum_i x_i u(x_i)$ and $\langle u^2
\rangle = \sum_i x_i u(x_i)^2$. Due to the definiteness of the sign
of $\dot U$, Lyapunov's theorem holds and, in particular, the
asymptotic state of the differential system is in the manifold where
$\dot U=0$, i.e. where $\langle u \rangle^2 =\langle u^2 \rangle $.
This condition, along with Eq. (\ref{norm}), is met if $N_0$ among
the $N$ population fractions  are equal to $N_0^{-1}$ ($1\le N_0\le
N$), while the remaining $N-N_0$ fractions are equal to zero. As
advanced from the analysis of Eq. (\ref{quot}), the asymptotic state
consists of a number of languages with identical non-vanishing
populations, while all the other languages are absent.

Note that the homogeneous stationary state where $x_i^* =N^{-1}$ for
all $i$, obtained in Section \ref{diff} for the case without
switching, is also found among the equilibrium solutions of the
system without mutation. Here, however, this equilibrium solution
corresponds to an extremely special initial condition, where all the
initial populations are identical. As a matter of fact, all the
equilibria with $N_0 >1$, where more than one language survives
asymptotically, are rather special, since they require that two or
more populations are initially identical. These equilibria would not
be robust under the effects of fluctuations in the initial
condition, or of noise during the evolution. In the most generic
case, on the other hand, there is a single maximal population. The
above analysis shows that this population will grow to the expense
of the others. In the absence of mutation, thus, the system
generically approaches a state where only one language survives.
This dominant language accumulates the whole population.

\section{Stationary states with mutation and switching} \label{stati}

Under the combined effects of mutation and switching, it is expected
that the system approaches a state where the population is
distributed among languages in a way that interpolates between the
stationary solutions discussed separately for each mechanism in
Sections \ref{diff} and \ref{repl}. With the notation introduced
there, Eq. (\ref{fund}) can be rewritten as
\begin{equation} \label{joint}
\dot x_i = \rho x_i [ \langle u \rangle -u(x_i) ] + \frac{\mu}{L}
\sum_{j=1}^N D_{ij} x_j  .
\end{equation}
Stationary solutions are given by equating the right-hand side to
zero for every $i=1,2,\dots,N$, and solving for $x_i$. It should be
clear by now that a particular solution is the homogeneous
distribution $x_i^*=N^{-1}$ for all $i$. Numerical simulations of
agent-based bit-string language models \cite{wich} confirm the
presence of this stationary state  of homogeneity for large mutation
rates. For small $\mu$, on the other hand, they suggest that there
is a stationary solution with a dominant language, which accumulates
most of the population, together with several less populated
languages. In terms of the results of Sections \ref{diff} and
\ref{repl}, this solution should be interpreted as the consequence
of the interplay between the replicator dynamics of switching, which
concentrates population in a single language, and mutation, which
redistributes part of the population among ``dialects'' around the
dominant language.

Generally, a joint Lyapunov functional for replicator dynamics and
diffusion does not exist, so that a global stability analysis of Eq.
(\ref{joint}) is not possible along the lines used in Section
\ref{repl}. To my knowledge, the only exception is the case of
linear unfitness, $u(x)=1-x$, where the functional
\begin{equation}
V({\bf x}) =\frac{\rho}{4} \sum_{i,j=1}^N x_i^2x_j^2-\frac{\rho}{3}
\sum_{i=1}^N x_i^3+\frac{\mu}{2L} \sum_{i,j=1}^N D_{ij}x_ix_j,
\end{equation}
satisfies $\dot x_i = \partial V/\partial x_i$, so that $\dot V \ge
0$.

In any case, the local stability of the homogeneous state
$x_i^*=N^{-1}$ can be analyzed by standard linearization of Eq.
(\ref{joint}). From this analysis it turns out that the homogeneous
state is a stable equilibrium above the critical mutation rate
\begin{equation} \label{mu1}
\mu_1 = \frac{\rho}{2N}  | u' (N^{-1} ) | ,
\end{equation}
where $u'(x)$ is the derivative of the unfitness function. Note
that, unless $u(x)$ has a singularity at $x=0$, the critical
mutation rate is very small if the number of languages $N$ is large.
This seems to strongly disagree with numerical results:\cite{wich}
with $\rho=1$, $L=8$ ($N=256$), and $u(x)=1-x^2$, agent-based
simulations indicate that homogeneity is asymptotically approached
for $\mu \gtrsim 0.14$, while the above equation predicts $\mu_1
\sim 10^{-4}$. As I show below, however, this discrepancy is
fallacious.

Performing a linear stability analysis for the equilibrium state
where the population is concentrated around a dominant language,
requires first to explicitly find the corresponding stationary
solution $x_i^*$, which turns out not to be trivial at all. The
equilibrium equations derived from Eq. (\ref{joint}) couple
nonlinearly, through the average $\langle u \rangle$,  all the
population fractions $x_i^*$. An approximate solution can however be
found in the limit of small mutation rates, $\mu \ll 1$, when
essentially all the population speaks the dominant language.
Assuming that the equilibrium population fraction $x_d^*$ of a
language at Hamming distance $d$ from the dominant one is of order
$\mu^d$, and keeping only the most significant terms in powers of
$\mu$, yields
\begin{equation} \label{peak}
x_d^* = d! \left( \frac{\mu}{\rho L}\right)^d x_0^*.
\end{equation}
The population fraction of the dominant language, $x_0^*$, can be
obtained from the normalization condition and, within the same
approximation order, is $x_0^* =1-\mu/\rho$. Note that this solution
for small $\mu$ is independent of the form of $u(x)$. This is in
agreement with numerical simulations for very small mutation rates,
which have found no sensible dependence on the unfitness
function.\cite{wich} Note also that the relevant small parameter in
this approximation is the ratio $\mu/\rho$. Linear stability
analysis of the solution (\ref{peak}) shows that, as expected, it is
stable for small $\mu$. An instability is predicted for
$\mu_2=\rho$, but this critical value of the mutation rate is
outside the validity range of the approximation. Comparing with Eq.
(\ref{mu1}), however, this result suggests that there may be an
interval in $\mu$, just above $\mu_1$, where {\it both} the
homogeneous state and the dominance state are stable. Such behaviour
would be consistent with known features in models of language
learning with mutations.\cite{koma} In the following section, I
support this conjecture from a different approach, and validate the
result by means of numerical resolution of Eq. (\ref{joint}).

\section{Bistability of dominance and homogeneity}

In order to progress beyond the limit of small mutation rates, the
solution to Eq. (\ref{joint}) must be approached from a different
perspective. I thus focus the attention on the evolution of the
maximal population fraction all over the system, $x_{\max}$, which I
assume is reached at a single language. Furthermore, invoking a kind
of mean-field approximation, I suppose that the population fractions
of  the remaining $N-1$ languages are mutually identical. Due to
normalization, their value is $(1-x_{\max})/(N-1)$. Replacing in Eq.
(\ref{joint}) yields an evolution equation for $x_{\max}$:
\begin{equation}  \label{xmax}
\dot x_{\max} = \rho x_{\max}(1-x_{\max})\left[ u
\left(\frac{1-x_{\max}}{N-1} \right) -u(x_{\max}) \right]-\mu\frac{N
x_{\max}-1}{N-1} .
\end{equation}

Encouragingly, $x_{\max}^* = N^{-1}$ is an equilibrium solution of
Eq. (\ref{xmax}), for any form of $u(x)$. For this solution, the
population fractions of all the other languages are also $N^{-1}$,
so that Eq. (\ref{xmax}) correctly predicts the existence of a
stationary homogeneous state. On the other hand, its stability
threshold is predicted by linearization of Eq. (\ref{xmax}) at
$\tilde \mu_1 = \rho(N-1)| u' (N^{-1} ) |/N^2 $ which, for large
$N$, differs from the critical mutation rate given in Eq.
(\ref{mu1}) by a factor of two. This is however understandable,
taking into account that not any deviation from homogeneity is
compatible with the present approximation, so that not all
eigenvalues of the original problem are at work in Eq. (\ref{xmax})
to bring the system to equilibrium. Nevertheless, the dependence of
$\tilde \mu_1$ and $\mu_1$ on $N$ is the same and, in particular,
$\tilde \mu_1$ tends to zero for large $N$.

Note also that, for $\mu=0$, $x_{\max}^* = 1$ is always an
equilibrium solution. It turns out to be stable for any decreasing
function $u(x)$. This solution corresponds to the dominance state in
the absence of mutation, where all the population speaks the same
language.

The existence and stability of other equilibrium solutions for Eq.
(\ref{xmax}) depend on the form of the unfitness $u(x)$. For the
sake of concreteness, I discuss the problem in the limit of large
$N$, where the equation for the equilibria of Eq. (\ref{xmax}) takes
the simpler form
\begin{equation}  \label{xmaxeq}
0 = \rho x_{\max}^*(1-x_{\max}^*)\left[1 -u(x_{\max}^*) \right]-\mu
x_{\max}^*.
\end{equation}
Here, I have assumed that $u(0)=1$. The homogeneous state is now
given by the trivial solution $x_{\max}^* = 0$. It can be easily
realized that, if $u(x)$ decreases with $x$ and $\mu$ is positive
but not too large, two additional solutions exist in the interval
$(0,1)$. The larger solution is a stable equilibrium of Eq.
(\ref{xmax}), and tends to $x_{\max}^* = 1$ for $\mu \to 0$, thus
corresponding to the dominance state. The lower solution is
unstable, and approaches zero in the same limit. As the mutation
rate grows, these two solutions approach each other, and eventually
collide and disappear through a tangent bifurcation at a critical
mutation rate $\mu_2$.

\begin{figure}[h]
\centerline{\psfig{file=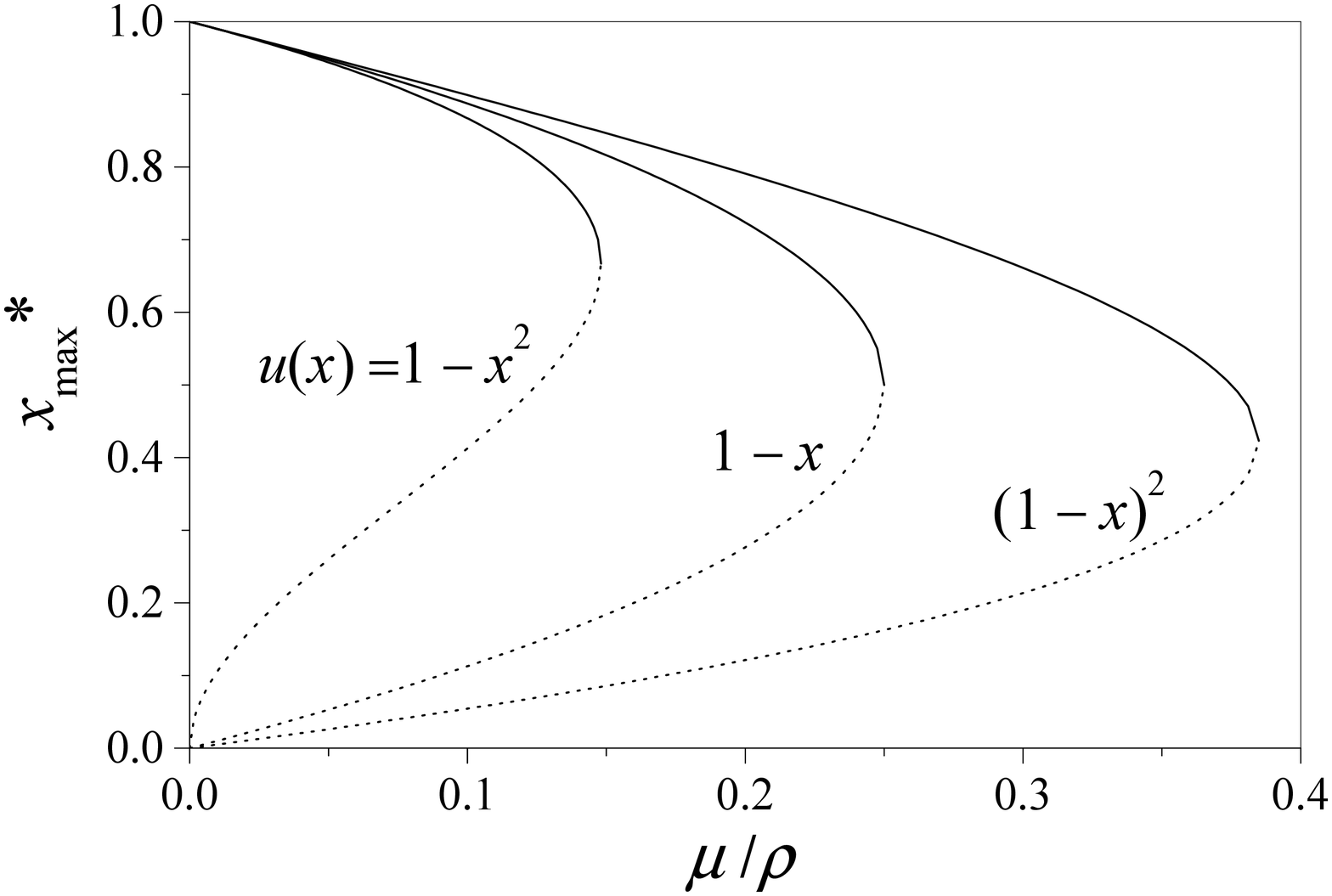,width=12cm}}  \vspace*{8pt}
\caption{The two solutions of Eq. (\ref{xmaxeq}) in the interval
$(0,1)$ as functions of the normalized mutation rate $\mu/\rho$, for
three forms of the unfitness function $u(x)$. Full lines stand for
the stable equilibria of Eq. (\ref{xmax}), in the limit $N\to
\infty$, corresponding to the dominance state. Dotted lines indicate
unstable equilibria. \label{f1}}
\end{figure}

The scenario is illustrated in Fig. \ref{f1} for three forms of the
unfitness $u(x)$. The curves show $x_{\max}^*$ as a function of the
(normalized) mutation rate $\mu/\rho$, corresponding to the
dominance state (full line) and the unstable solution (dotted line).
For these low-degree polynomial forms of $u(x)$ the critical point
of the tangent bifurcation can be exactly calculated. For
$u(x)=(1-x)^2$, $1-x$, and $(1-x)^2$, the respective critical
mutation rates are $\mu_2 = 4 \rho/27 \approx 0.148 \rho$, $\rho/4$,
and $2\rho /3\sqrt 3\approx 0.385 \rho$. The first value is in very
good agreement with the critical point reported from agent-based
numerical simulations ($\mu=0.14$ for $\rho=1$ and
$N=256$),\cite{wich} and makes it possible to identify the
transition detected numerically as the tangent bifurcation at
$\mu_2$.

As advanced at the end of Section \ref{stati}, the present results
predict that, for mutation rates satisfying $\mu_1<\mu<\mu_2$, both
dominance and homogeneity are stable states for the distribution of
population among languages. In other words, two stable solutions
coexist and can be asymptotically approached during the evolution.
The asymptotic state is selected by the initial condition: in Eq.
(\ref{xmax}), the attraction basins of the two solutions are
separated by the intermediate unstable state. This dependence on the
initial condition has also been noticed in agent-based
simulations.\cite{life,sol}

\begin{figure}[h]
\centerline{\psfig{file=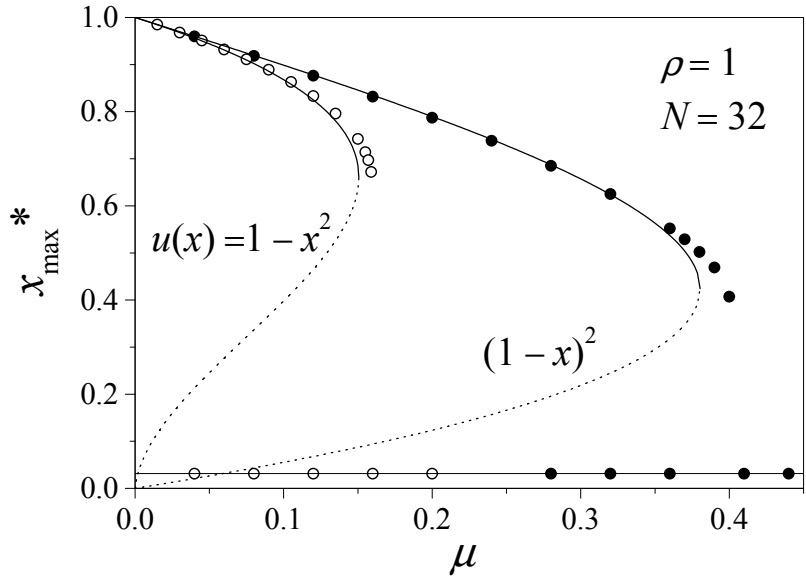,width=12cm}}  \vspace*{8pt}
\caption{Full and dotted curves respectively stand for the stable
and unstable equilibrium solutions to Eq. (\ref{xmax}), as functions
of the mutation rate, for $\rho=1$ and $N=32$ ($L=5$) and two forms
of the unfitness $u(x)$. Dots correspond to long-time measurements
of $x_{\max}$ from numerical resolution of Eq. (\ref{joint}) for
each form of $u(x)$. \label{f2}}
\end{figure}

In spite of the agreement with simulations, Eq. (\ref{xmax}) remains
the outcome of a rather rough assumption on the population
distribution over languages. It is therefore worthwhile to compare
its predictions with results from the numerical resolution of Eq.
(\ref{joint}). This is also an opportunity to consider relatively
small values of $N$, which have been disregarded in Eq.
(\ref{xmaxeq}). Curves in Fig. \ref{f2} show the equilibria of Eq.
(\ref{xmax}) as functions of the mutation rate $\mu$, for $L=5$
($N=32$), $\rho=1$, and two forms of the unfitness $u(x)$. Dots
stand for the maximal population fraction obtained, at long times,
from the numerical resolution of Eq. (\ref{joint}) with each
function $u(x)$. Two kinds of initial conditions have been
considered. In one of them, the initial population fractions are
chosen at random, all of them close to the homogeneous state, and
satisfying the normalization (\ref{norm}). In the other initial
condition, one of the population fractions equals unity, and all the
other are zero. As expected, the former are found to asymptotically
approach the value $x_{\max}^* = N^{-1}\approx 0.031$. When the
initial population is concentrated in a single language, on the
other hand,  $x_{\max}^*$ is relatively large for mutation rates
below the tangent bifurcation, and drops to $N^{-1}$ above it. The
agreement between the results from Eqs. (\ref{joint}) and
(\ref{xmax}) is excellent for small values of $\mu$ and, perhaps not
unexpectedly, worsens as the critical point $\mu_2$ is approached.
However, the overall pictures are qualitatively identical.

\section{Conclusion}

In this paper, I have presented a formulation of bit-string models
of language evolution based on differential equations for the
fraction of the population speaking each language. The formulation
highlights the fact that these models conceive language evolution as
combining a diffusion mechanism, given by mutation between similar
languages, with replicator dynamics for language switching, when an
individual adopts the language of a randomly selected member of the
population. The combination of replicator dynamics and diffusion is
not new in the literature of biological evolution models. In
contrast with bit-string language models, however, these
applications to biological evolution often assume that the fitness
of individuals of each species is independent of the population,
though it varies between species.\cite{kimura,volken} In bit-string
models, on the other hand, the dependence of the unfitness on the
population fraction is essential to represent preference for more
widespread languages.

The decrease of the unfitness of a language when its population
grows is the key ingredient which shapes the behaviour of bit-string
models. In particular, the occurrence of a tangent bifurcation where
the dominance state --with most of the population speaking the same
language-- disappears, and the existence of a parameter region where
dominance and homogeneity coexist and are stable, are direct
consequences of such dependence. This bistability opens the
possibility that, in a system of languages divided into
weakly-interacting domains, where inter-domain switching is much
less likely than intra-domain transitions, some domains converge to
the dominance state while others approach homogeneity, even when the
evolution parameters are identical all over the system.

It is worth remarking that a  dominance-homogeneity transition is
known to happen in genetic space within replication-mutation models
of molecular evolution, specifically, in Eigen's
model.\cite{eigen,volken} Though the involved critical phenomenon is
mathematically different from the tangent bifurcation disclosed
here, they are qualitatively much the same from the viewpoint of the
competing balance between the basic mechanisms at work.

It would be interesting, by inspiration from mathematical studies of
biological evolution, to add bit-string models with some
non-uniformity in the individual properties of languages, for
instance, assigning different parameters to the unfitness function
of each language. In fact, it may well be that the preference for a
very widespread language is inhibited by intrinsic difficulties to
acquire it. {\it Vice versa}, a language whose acquisition is
perceivably easier could be chosen as a common communication means
by populations speaking more widespread but more intricate
languages. Do you imagine the whole of  mankind eventually deciding
to speak Mandarin and its dialects?\cite{sol} Think also of Mark
Twain's ``The horrors of the German language.''\cite{pinker}

Let me finally point out that comparison of numerical results from
bit-string models and empirical data of language statistics
--specifically, the distribution of language sizes, and the
abundance within language families-- has been focused on transient
stages of the evolutionary process. In agreement with this, though
not analyzed in detail here, the present study shows that asymptotic
states in bit-string models are not a good representation of
empirical observations. Transient effects in the present formulation
could be addressed by numerical resolution of Eqs. (\ref{fund}), but
such analysis would deserve a separate presentation.

\end{document}